# Simple guidelines to predict self-phase modulation patterns

CHRISTOPHE FINOT, [1,*] F. CHAUSSARD,[1] SONIA BOSCOLO [2]

[1]*Laboratoire Interdisciplinaire Carnot de Bourgogne, UMR 6303 Université de Bourgogne-Franche-Comté – CNRS, 9 avenue Alain Savary, Dijon, FRANCE*
[2]*Aston Institute of Photonic Technologies, School of Engineering and Applied Science, Aston University, Birmingham B4 7ET, United Kingdom*
*Corresponding author: christophe.finot@u-bourgogne.fr*

**We present a simple approach to predict the main features of optical spectra affected by self-phase modulation (SPM), which is based on regarding the spectrum modification as an interference effect. A two-wave interference model is found sufficient to describe the SPM-broadened spectra of initially transform-limited or up-chirped pulses, whereas a third wave should be included in the model for initially down-chirped pulses. Simple analytical formulae are derived, which accurately predict the positions of the outermost peaks of the spectra.**

## 1. INTRODUCTION

Self-phase modulation (SPM) is one of those very fascinating effects discovered in the early days of nonlinear optics [1] and among the first examples of nonlinear optical phenomena to which novices of the field are usually exposed [2, 3]. It refers to the phenomenon by which an intense optical beam propagating in a Kerr medium induces through the nonlinearity of the medium a modulation of its phase that is proportional to its own intensity profile. Fiber optics is a convenient testbed for the experimental study of nonlinear optical phenomena owing to the possibility for a propagating beam to undergo large amounts sof nonlinearity without suffering from spatial modifications or thermal effects [4]. The earliest observation of SPM in optical fibers was made in 1978 [5], and the phenomenon has been extensively studied since then. For an input pulsed beam, the time-dependent phase change induced by SPM is associated with a modification of the optical spectrum, which depends on the frequency modulation (chirp) of the pulse electric field. If the pulse is initially Fourier-transform-limited or up-chirped SPM leads to spectral broadening, whereas an initially down-chirped pulse is spectrally compressed by the effects of SPM [5, 6]. For strong SPM, the optical spectrum can exhibit strong oscillations.

SPM is often regarded as being harmful for optical communication systems or high-power pulse generation [7]. However, when conveniently managed, SPM can be a precious ingredient that has stimulated a tremendous amount of applications, including the generation of ultra-short pulses [8, 9], the compensation of dispersion through solitonic effects [4, 10], the generation of wavelength-multiplexed sources [11], the frequency shifting of ultra-short pulse sources [12], the contrast enhancement of ultra-short pulses [13], several applications in ultrafast optical signal processing [14, 15], and the characterization of optical pulses [16, 17] and waveguides [18].

In this paper, by extending an approach initially developed in the context of sinusoidally phase modulated continuous waves [19], we present a novel theoretical treatment of SPM based on a spectral interference model. The typical oscillatory character of optical spectra affected by SPM indeed originates from strong excursions of the instantaneous frequency, so that in general there are contributions from different times to the Fourier integral for a given frequency component. Depending on the exact frequency, these contributions may constructively add up or cancel each other [1]. After introducing the situation under investigation, we present the intuitive but accurate description of the SPM spectral patterns that is afforded by our proposed approach. Simple analytical formulae are derived, which accurately predict the positions of the outermost peaks of the spectra. In the last section, we discuss both qualitatively and quantitatively the different scenarios that are observed depending on the initial chirp of the pulses.

## 2. PRINCIPLE AND SITUATION UNDER INVESTIGATION

Pulse propagation in a single-mode optical fiber is modeled with the standard nonlinear Schrödinger equation [4]:

$$i \frac{\partial \psi}{\partial z} = \frac{1}{2} \beta_2 \frac{\partial^2 \psi}{\partial t^2} - \gamma |\psi|^2 \psi, \qquad (1)$$

where $\psi(z,t)$ is the complex envelope of the pulse electric field, $z$ is the propagation coordinate, $t$ is the retarded time, $\beta_2$ is the group-velocity dispersion parameter, and $\gamma$ is the Kerr-nonlinearity coefficient accounting for both the nonlinear refractive index $n_2$ and the fiber's effective cross-sectional area. We consider an initial pulse having the waveform $\psi(t,0) = \sqrt{P_0}\sqrt{I(t)} \exp\left(-C\, t^2/(2T_0^2)\right)$, where $I(t)$ is the normalized temporal intensity profile of the pulse, $P_0$ is the peak power, $T_0$ is a characteristic time associated with the pulse, and $C$ is a chirp coefficient, which can be positive or negative. When the effects of Kerr nonlinearity are considered over those distances and power levels such that we can neglect dispersion (i.e. when the condition $1/(\gamma P_0) \ll T_0^2 / \beta_2$ is satisfied), the solution to Eq. (1) is simply

$$\psi(z,t) = \psi(0,t) \exp\left(i\, \varphi_{NL}(z,t)\right), \qquad (2)$$

where $\varphi_{NL}(z,t) = B\, I(t)$ is the SPM-induced nonlinear phase shift. Here, $B = \gamma P_0\, z$ is the maximum phase shift that occurs at the pulse center located at $t = 0$ and is widely known as the *B*-integral. In the presence of loss or gain, the physical propagated length $z$ is replaced with an effective length defined as $L_{eff} = [1 - \exp(-\alpha z)] / \alpha$, where $\alpha > 0$ (< 0) accounts for loss (gain). The temporally varying nonlinear phase implies a time dependence of the instantaneous optical frequency $\delta\omega_{NL}(t) = -d\varphi_{NL}/dt$, which in turn translates into changes in the pulse spectrum. The shape of the spectrum $S(\omega)$ is obtained by taking the Fourier transform of Eq. (2):

$$S(\omega) = \left| \int_{-\infty}^{\infty} \psi(0,t) \exp(i B I(t)) \exp(i \omega t) \, dt \right|^2. \quad (3)$$

In general, the spectrum depends on the pulse shape and the initial chirp imposed on the pulse. In the general case, $S(\omega)$ cannot be calculated analytically. However, for large values of the *B*-integral, many important insights into $S(\omega)$ can be obtained using the method of stationary phase [20].

For an input pulse with a parabolic shape, the SPM-induced frequency chirp $\delta\omega_{NL}$ is a strictly monotonic function of time, so the pulse has a different instantaneous frequency at each point. In this case, no spectral interference occurs and, thus, the spectrum does not feature any oscillatory structure [21]. Conversely, for common transform-limited bell-shaped pulses, such as Gaussian or hyperbolic secant pulses, $\delta\omega_{NL}$ has a non-monotonic temporal variation; it reaches maximum positive and minimum negative values, $\omega_m$ and $\omega_{min}$, and approaches zero as $t$ becomes infinitely large. It has therefore been proposed to estimate the magnitude of SPM-induced spectral broadening by simply calculating $\Delta\omega = \omega_m - \omega_{min}$. A more accurate measure of spectral broadening is provided by the root-mean-square (rms) spectral width $\omega_{rms}$ [22]. The typical time dependence of $\delta\omega_{NL}$ for these bell-like pulse waveforms also means that for $\omega_{min} < \delta\omega_{NL} < \omega_m$ the same chirp occurs at two values of $t$, showing that the pulse has the same instantaneous frequency at two distinct points. Qualitatively speaking, these two points represent two waves of the same frequency but different phases that can interfere constructively or destructively depending on their relative phase difference. The characteristic oscillatory structure in the pulse spectrum is a result of such interference [1]. Mathematically, the Fourier integral in Eq. (3) gets dominant contributions at the two values of $t$ at which the chirp is the same. These contributions may add up in phase or out of phase. Indeed, it is possible to use the stationary phase method to obtain an approximate expression of $S(\omega)$ that is valid for large values of $B$ [20]. Note that for initially chirped pulses, in general the same chirp may occur at more than two values of $t$. The approach that we propose here is to evaluate the spectrum affected by SPM by calculating the intensity in the interference pattern associated with the different instants in the pulse that have the same instantaneous frequency. A measure of the spectrum extent is then provided by the position of the outermost peak of intensity.

## 3. SPM PATTERN PREDICTION OF INITIALLY TRANSFORM-LIMITED PULSE

### A. Analysis of the spectrum of a Gaussian pulse

*1. Analysis*

In order to introduce the basis of our discussion, we first consider the case of the widely used Fourier-transform-limited Gaussian pulse with the temporal intensity profile: $I(t) = G(t) = \exp(-(t/T_0)^2)$, where $T_0$ is the half-width at $1/e$ intensity point. Given the evenness of the intensity function and the oddness of the chirp function, we can restrict our study to the positive times only. The chirp $\delta\omega(t) = \delta\omega_{NL}(t)$ for such a pulse [Fig. 1(b)] is

$$\delta\omega(t) = \frac{B}{T_0} H_1\left(\frac{t}{T_0}\right) G(t), \quad (4)$$

where $H_n$ is the $n^{th}$-order Hermite polynomial. The extrema of the chirp function can be easily found by solving the equation

$$\frac{d\delta\omega(t)}{dt} = -\frac{B}{T_0^2} H_2\left(\frac{t}{T_0}\right) G(t) = 0, \quad (5)$$

which yields the following maximum value of $\delta\omega$ on the positive time $t_m$:

$$\begin{cases} t_m = \dfrac{T_0}{\sqrt{2}} \\ \omega_m = \dfrac{B}{t_m} \exp\left(-\dfrac{1}{2}\right) \end{cases}. \quad (6)$$

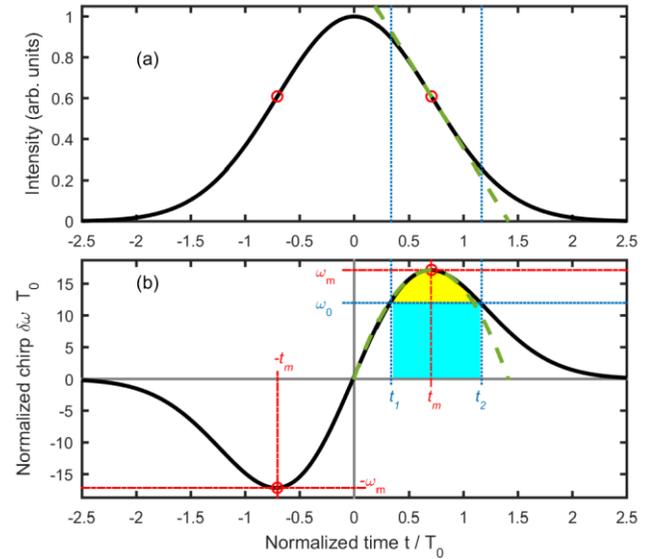

Fig. 1. Temporal intensity and chirp profiles for a transform-limited Gaussian pulse after undergoing SPM at a *B*-integral of 20 rad. The extreme points of the chirp function and the solutions of the equation $\delta\omega = \omega_0$ are indicated with red and blue lines, respectively. The cyan and yellow regions represent the phase difference $\Delta\varphi_D$ and the opposite of the total phase offset $\Delta\varphi_T$, respectively. The approximation to the chirp function based on Eq. (15) and the corresponding approximation to the pulse profile are plotted with dashed green lines.

As can be seen in Fig. 1(b), a given instantaneous frequency $\omega_0$ between 0 and $\omega_m$ is observed at two different instants, $t_1$ and $t_2$ (with $t_2 > t_1$), whose separation $\Delta t = t_2 - t_1$ is a decreasing function of $\omega_0$. These instants are the two roots of the equation $\delta\omega(t) = \omega_0$:

$$\begin{cases} t_1 = t_m \sqrt{-W_0\left(-\left(\frac{\omega_0}{\omega_m}\right)^2 e\right)} \\ t_2 = t_m \sqrt{-W_{-1}\left(-\left(\frac{\omega_0}{\omega_m}\right)^2 e\right)} \end{cases}, \quad (7)$$

where $W_n$ is the Lambert $W$ function, with $n$ = 0, −1 for the main and second branch, respectively. The total phase difference $\Delta\varphi_T$ between $t_1$ and $t_2$ is the result of two contributions. The first contribution $\Delta\varphi_D$ is due to the time delay $\Delta t$:

$$\Delta\varphi_D(\omega_0) = \omega_0\, \Delta t. \quad (8)$$

This quantity can be represented graphically in Fig. 1(b) as the area of the cyan rectangle, and is a non-monotonic function of $\omega_0$ (Fig. 2). The second contribution $\Delta\varphi_{NL}$ to the total phase offset arises from SPM:

$$\Delta\varphi_{NL}(\omega_0) = \varphi_{NL}(t_2) - \varphi_{NL}(t_1) = B[G(t_2) - G(t_1)], \quad (9)$$

and can be represented as the opposite of the area under the curve $\delta\omega(t)$ between $t = t_1$ and $t = t_2$ in Fig. 1(b). It is apparent that $\Delta\varphi_{NL}$ is a negative and monotonically increasing function of $\omega_0$ that has the maximum negative value of $-B$ at $\omega_0 = 0$ and is zero when $\omega_0 = \omega_m$. The total phase difference between $t_2$ and $t_1$ is therefore

$$\Delta\varphi_T(\omega_0) = \Delta\varphi_D + \Delta\varphi_{NL}. \quad (10)$$

It can be geometrically interpreted as the opposite of the area of the region in Fig. 1(b) that is bounded by the graph of $\delta\omega$ and the horizontal line $\delta\omega = \omega_0$ (yellow region), and it is a negative and monotonically increasing function of $\omega_0$ with the same extrema as $\Delta\varphi_{NL}$ (Fig. 2).

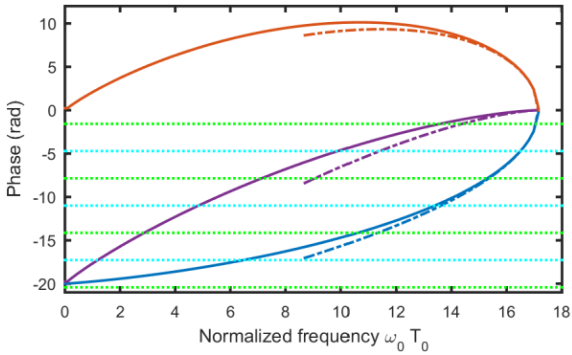

Fig. 2. Variation of the different phase differences between the two interfering times at the frequency $\omega_0$, $t_1$ and $t_2$, with $\omega_0$ for a transform-limited Gaussian pulse after undergoing SPM at $B$ = 20 rad. The phase differences $\Delta\varphi_D$, $\Delta\varphi_{NL}$ and $\Delta\varphi_T$ as calculated from Eqs. (8), (9) and (10) are represented by red, blue and purple solid curves, respectively. Their toy-model approximations $\Delta'\varphi_D$, $\Delta'\varphi_{NL}$ and $\Delta'\varphi_T$ based on Eq. (17) are represented by dash-dotted curves. The green and cyan horizontal lines represent the values of the total phase offset causing constructive and destructive interference, respectively [Eqs. (13) and (12)].

With the knowledge of the instants $t_1$ and $t_2$ and the total phase difference $\Delta\varphi_T$ between them, we can calculate the distribution of intensity $I(\omega_0)$ in the pattern associated with the interference between these two times:

$$I(\omega_0) = G(t_1) + G(t_2) + 2\sqrt{G(t_1)\, G(t_2)} \cos\left(\Delta\varphi_T(\omega_0) + \frac{\pi}{2}\right). \quad (11)$$

Equation (11) indicates that there is destructive interference causing intensity minima at frequencies for which

$$\Delta\varphi_T = (2\, m+1)\,\pi - \frac{\pi}{2}, \quad (12)$$

with $m$ being a negative integer number (Fig. 2, cyan lines). From Eq. (12), one may obtain the convenient relationship between the number of minima $k$ on one side of the spectrum and the value of $B$ derived by Cubeddu et al. in [20]: $B = (2k-1)\,\pi$. Intensity maxima or constructive interference occur when

$$\Delta\varphi_T = 2\, m\, \pi - \frac{\pi}{2} \quad (13)$$

at negative integer $m$. In particular, the position $\omega_M$ of the outermost peak of intensity corresponds to $m = 0$ in Eq. (13):

$$\Delta\varphi_T(\omega_M) = -\frac{\pi}{2} \quad (14)$$

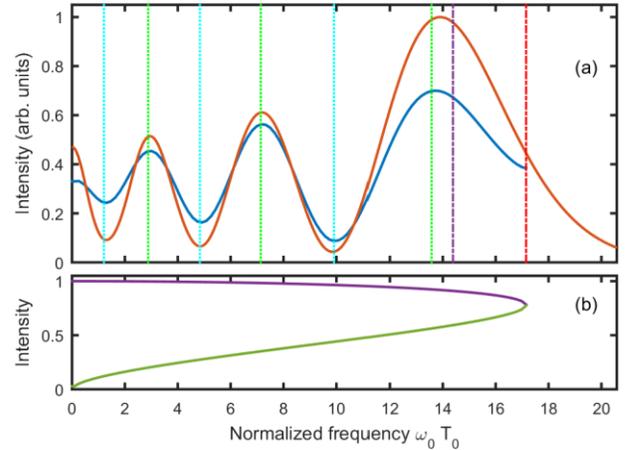

Fig. 3. (a) Spectral intensity distribution $I(\omega_0)$ calculated from Eq. (11) (blue solid curve) compared to the spectrum $S(\omega)$ obtained from Eq. (3) (red solid curve) for a transform-limited Gaussian pulse after undergoing SPM at $B$ = 20 rad. The green and cyan dotted lines represent the frequencies at which constructive and destructive interference occurs, as predicted by Eqs. (13) and (12), respectively. The red and purple vertical dash-dotted lines represent the maximum frequency $\omega_m$ from Eq. (6) and the toy-model prediction for the position of the outermost spectral peak $\omega'_M$ [Eq. (20)]. (b) Evolutions of the intensities of the two interfering waves at the frequency $\omega_0$, $I_1 = G(t_1)$ and $I_2 = G(t_2)$, with $\omega_0$ (purple and green curves, respectively).

It is worth noting that $\Delta\varphi_T$ being not a linear function of $\omega_0$, the frequencies for which constructive and destructive interference occurs are not equally spaced, and the outermost intensity peak is the widest [23]. Figure 3 compares the spectral intensity distribution $I(\omega_0)$ given by Eq. (11) with the exact spectrum $S(\omega)$ numerically calculated from Eq. (3) at a level of SPM corresponding to $B$ = 20 rad, and apparently indicates excellent qualitative agreement between the two spectral patterns. The positions of the intensity extrema predicted by Eqs. (12) and (13) are in quantitative agreement with the actual positions. The evolutions of the intensities of the two interfering waves at the instantaneous frequency $\omega_0$, $I_1 = G(t_1)$ and $I_2 = G(t_2)$, with $\omega_0$ that are plotted in Fig. 2(b) highlight that the lower $\omega_0$, the higher the intensity difference between these two waves. This brings about a decreasing contrast in the interference pattern $I(\omega_0)$ with decreasing $\omega_0$. It is also noteworthy that our model is intrinsically unable to predict the outer decreasing wings of the spectrum.

We have also compared the evolutions of $I(\omega_0)$ and $S(\omega)$ with the level of SPM accumulated in the fiber. The results shown in Fig. 4(a) confirm both the qualitative agreement between the spectral pattern predicted by Eq. (11) and the numerical evaluation of the spectrum from Eq. (3), and the accuracy of the analytical prediction for the positions of the spectrum's extrema. It is also apparent from Fig. 4(a) that the maximum instantaneous frequency excursion contained in the spectrum $\omega_m$ or the rms spectral width $\omega_{rms} = \omega_{rms,0} [1 + (0.877 B)^2]^{1/2}$ derived by Pinault et al. in [22] ($\omega_{rms,0}$ is the initial rms spectral width of the pulse) may not be the most intuitive quantities to describe the spectrum's expansion generated by SPM and, especially, the position of the outermost spectral peak that is very relevant for many recent applications of SPM [12, 13]. Our approach is also able to describe the evolution of the spectral intensity at the central frequency $\omega_0 = 0$, for which $\Delta\varphi_T = -B$, with the B-integral. A simple sinusoidal function can indeed qualitatively explain the resulting pattern: $I(B, \omega_0=0) \propto 1 + \sin(B)/2$ [Fig. 4(b)]. This can be of interest for nonlinear signal processing applications based on optical band-pass filtering at the central frequency [14, 15].

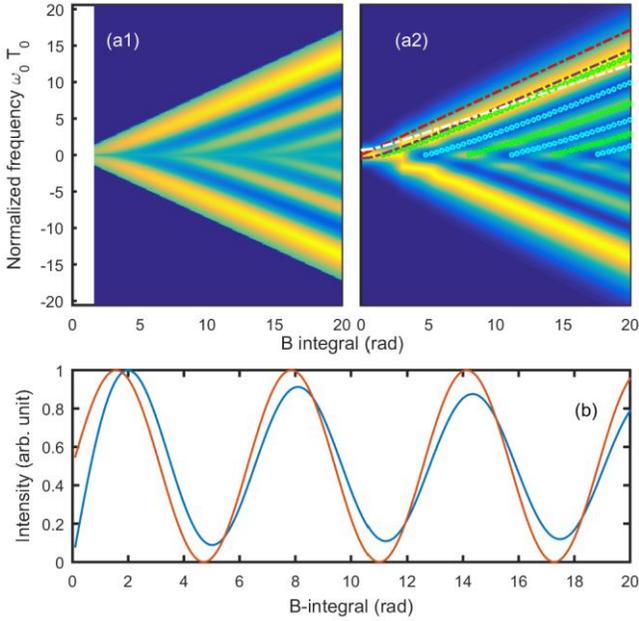

Fig. 4. (a) Evolution of the spectrum of a transform-limited Gaussian pulse with the B-integral accumulated in the fiber. The spectral intensity distribution $I(\omega)$ calculated from Eq. (11) (panel 1) is compared to the spectrum $S(\omega)$ obtained from Eq. (3) (panel 2). The evolutions of the rms spectral width $\omega_{rms}$, the position of the outermost spectral peak $\omega_M$ predicted by Eq. (14), and the value predicted by the toy model $\omega'_M$ [Eq. (20)] are plotted with red, white and dark blue dash-dotted lines, respectively. The green and cyan dotted open circles represent the frequencies at which constructive and destructive interference occurs, as predicted by Eqs. (13) and (12), respectively. (b) Evolution of the spectral intensity at the central frequency with the accumulated B-integral as obtained by numerical evaluation of Eq. (3) (red curve) and by the relationship: $I(B, \omega_0=0) \propto 1 + \sin(B)/2$ stemming from Eq. (11) (blue curve).

*2. Toy model*

One point of the analysis so far that could be slightly inconvenient is the use of the Lambert W function to determine the interfering instants $t_1$ and $t_2$. While such a function is nowadays widely implemented in professional scientific software, it prevents the development of a fully analytical approach to the problem. Simplified but fully tractable expressions giving the main features of the spectral interference process can be obtained by expanding the chirp function $\delta\omega(t)$ in Taylor series about $t = t_m$ to second order:

$$\delta\omega(t) \simeq \omega_m + A\left(t - t_m\right)^2,$$
$$A = \frac{1}{2}\left.\frac{d^2\delta\omega}{dt^2}\right|_{t_m} = -2\frac{\omega_m}{T_0^2} = -\frac{\omega_m}{t_m^2}. \quad (15)$$

As we can see in Fig. 1(b) (dashed green curve), this expansion provides a rather good approximation to the actual function in the vicinity of $\omega_m$. Within this approximation, the interfering instants $t'_1$ and $t'_2$ are simply the roots of a quadratic equation, $\delta\omega(t') = \omega_0$, hence they are separated by

$$\Delta T' = t'_2 - t'_1 = 2 t_m \sqrt{\frac{\omega_m - \omega_0}{\omega_m}}, \quad (16)$$

and the linear and nonlinear phase differences between them are

$$\begin{cases} \Delta\varphi'_D = \omega_0 \Delta T' \\ \Delta\varphi'_{NL} = -\omega_m \Delta T' \end{cases}. \quad (17)$$

Here, the nonlinear phase contribution $\Delta\varphi_{NL}$ has been obtained by using a first-order Taylor expansion of the pulse intensity profile $I(t)$ about $t_m$: $I(t) = I(t_m) - \omega_m(t - t_m)/B$ [Fig. 1(a), green dashed curve]. The predictions from Eq. (17) show fairly good agreement with the results obtained from Eqs. (8), (9) and (10) as long the instantaneous frequency remains close to its maximum value $\omega_m$ (Fig. 2, dash-dotted curves). Accordingly, Eq. (14) giving the position of the outermost spectral peak simplifies to

$$\Delta\varphi'_T = \left(\omega'_M - \omega_m\right)\Delta T' = -\frac{\pi}{2}. \quad (18)$$

This equation can be readily solved to yield

$$\omega'_M = \omega_m - \frac{\pi^{2/3}}{2}\frac{\omega_m^{1/3}}{T_0^{2/3}}. \quad (19)$$

For a Gaussian pulse Eq. (19) takes the form

$$\omega'_M = \omega_m\left(1 - \left(\frac{\pi}{4B}\right)^{\frac{2}{3}}\exp\left(\frac{1}{3}\right)\right) \simeq \omega_m\left(1 - 1.84\, B^{-\frac{2}{3}}\right). \quad (20)$$

Therefore, our toy model explicitly predicts a B-dependent correction factor for the position of the outermost spectral peak relative to the maximum instantaneous frequency excursion. As can be seen in Figs. 3 and 4 (purple dashed curves), the prediction based on Eq. (20) is entirely plausible, thereby being more convenient for practical applications than $\omega_{rms}$ or $\omega_m$.

## B. Analysis of other pulse shapes

We have also used our proposed method to describe the SPM-broadened spectra of initially transform-limited pulses with hyperbolic secant, Lorentzian and super Gaussian temporal intensity profiles. In the case of a hyperbolic secant pulse with $I(t) = \text{sech}^2(t/T_0)$, the chirp is given by $\delta\omega(t) = 2 B I(t) \tanh(t/T_0)$, and attains a maximum value of $\delta\omega_m = 4 B 3^{-3/2} / T_0$ at $t_m = T_0 \text{atanh}(3^{-1/2})$. The toy model supplies the same expression for the position of the outermost spectral peak as that of Eq. (20). We note that for this pulse shape, an exact calculation of the optical spectrum affected by SPM is available [24]. However, this calculation remains quite technical and not fully intuitive.

The Lorentzian profile $I(t) = (1+(t/T_0)^2)^{-4}$ entails the chirp $\delta\omega(t) = 4 B (1+(t/T_0)^2)^{-3}/T_0$ with a maximum value of $\omega_m = 25 B / (54 t_m)$ at $t_m = T_0 5^{-1/2}$. The toy-model analysis leads to the following prediction of the position of the outermost spectral peak:

$$\omega'_M = \omega_m \left(1 - \frac{3^{5/3}}{10}\left(\frac{\pi}{B}\right)^{\frac{2}{3}}\right) \simeq \omega_m \left(1 - 1.34 B^{-\frac{2}{3}}\right) \quad (21)$$

The results obtained for the hyperbolic secant and Lorentzian pulse shapes are summarized in Fig. 5, and demonstrate the excellent ability of our approach to reproduce the overall shape and the extreme points of the spectrum. However, similarly to the case of a Gaussian pulse, the amplitude of the outermost peak is underestimated. This stems from the fact that the stationary-phase approximation ceases to be valid in the vicinity of $\omega_m$.

Figure 6 shows the results obtained for a super-Gaussian pulse given by $I(t) = \exp(-(t/T_0)^{2m})$ with $m = 3$. In this case, the chirp $\omega_m$ has a maximum point at $t_m = T_0(1-1/(2m))^{1/2m}$, which can be approximated to $T_0$ for large $m$. With this approximation, we get $\omega_m = 2 B m e^{-1} / T_0$. Thus, using the toy model, we obtain the following expression for the position of the outermost spectral peak:

$$\omega'_M = \omega_m \left(1 - \frac{1}{2}\left(\frac{\pi}{2B}\right)^{\frac{2}{3}} \exp\left(\frac{2}{3}\right)\right) \simeq \omega_m \left(1 - 1.32 B^{-\frac{2}{3}}\right). \quad (22)$$

We can see in Fig. 6 that our method can accurately predict the lateral extreme points of the spectrum. However, a significant discrepancy between the spectral intensity distribution $I(\omega)$ and the actual spectrum $S(\omega)$ can be observed at the low frequencies. In particular, $S(\omega)$ features a marked peak at $\omega = 0$, which cannot be reproduced by our approach. In fact, most of the pulse energy remains in the central peak because the SPM-induced chirp is nearly zero over the central region of the pulse as a consequence of the nearly uniform intensity of a super-Gaussian pulse for $|t| < T_0$. This central peak may make the first minimum from the center slightly fuzzy, hence care should be taken when applying the formula for the number of minima in the SPM-broadened spectrum derived by Cubeddu *et al.* to experimental data.

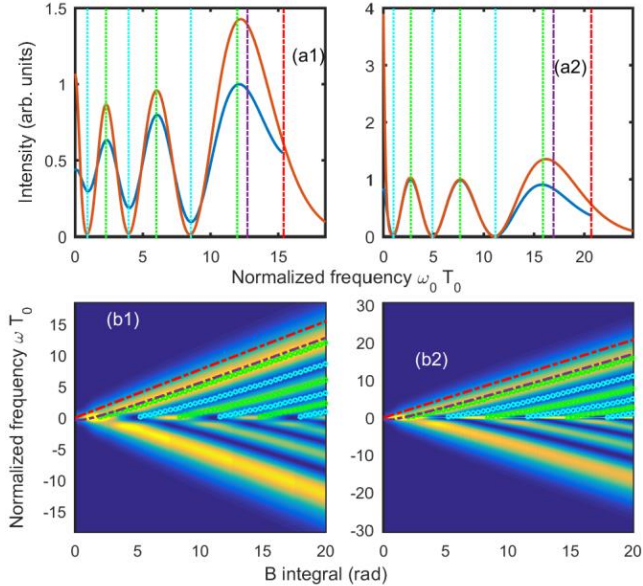

Fig. 5. SPM patterns of transform-limited hyperbolic secant (panels 1) and Lorentzian (panels 2) pulses. (a) Spectral intensity distribution $I(\omega)$ (blue solid curve) compared to the spectrum $S(\omega)$ obtained from Eq. (3) (red solid curve) for a $B$-integral accumulated in the fiber of 20 rad. (b) Evolution of the spectrum $S(\omega)$ with the $B$-integral. The red and purple vertical dash-dotted lines represent the maximum frequency $\omega_m$ and the toy-model prediction for the position of the outermost spectral peak $\omega'_M$ [Eqs. (20) and (21)]. The green and cyan dotted lines or open circles represent the frequencies at which constructive and destructive interference occurs, as predicted by Eqs. (13) and (12), respectively.

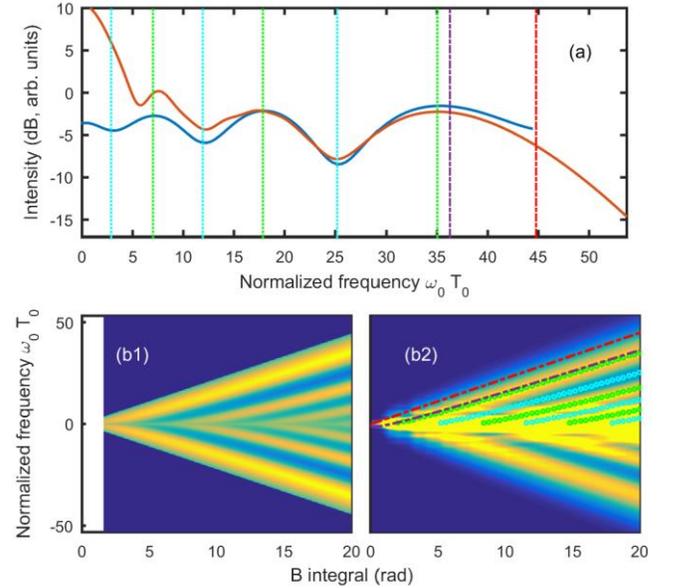

Fig. 6. SPM pattern of a transform-limited super-Gaussian pulse with $m = 3$. (a) Spectral intensity distribution $I(\omega)$ (blue solid curve) compared to the spectrum $S(\omega)$ obtained from Eq. (3) (red solid curve) for a $B$-integral accumulated in the fiber of 20 rad. (b) Evolutions of $I(\omega)$ (panel 1) and $S(\omega)$ (panel 2) with the $B$-integral. The red and purple vertical dash-dotted lines represent the maximum frequency $\omega_m$ and the toy-model prediction for the position of the outermost spectral peak $\omega'_M$ [Eq. (22)]. The green and cyan dotted lines or open circles represent the frequencies at which constructive and destructive interference occurs, as predicted by Eqs. (13) and (12), respectively.

# 4. SPM PATTERN PREDICTION OF LINEARLY CHIRPED INPUT PULSE

## A. General discussion on the appearance of spectral interference fringes

In this section, we discuss the effects of an initial linear chirp on the SPM-modified pulse spectrum. Such a chirp can be easily generated through propagation in a linear dispersive medium such as a pair of diffraction gratings, fiber Bragg gratings or a hollow core optical fiber. Linearly chirped pulses are widely used in chirped pulse amplification to mitigate undesirable SPM [7, 25]. However, the residual SPM can be sufficient to degrade the quality of the pulses [26]. We focus our discussion on an input Gaussian pulse, but the method can be extended to other bell-shaped waveforms [27]. Note that part of the discussion can also aid to qualitatively understand the initial stage of wave-breaking occurring in a nonlinear dispersive fiber, as described by Anderson *et al.* [28]. For a Gaussian pulse with an initial linear chirp ($C \neq 0$), the total chirp after undergoing SPM is given by

$$\delta\omega(t) = \frac{1}{T_0} H_1\left(\frac{t}{T_0}\right)\left(B\, G(t) + \frac{C}{2}\right). \quad (23)$$

Hence, the extreme points of the chirp $t_e$ verify the following equation:

$$H_2\left(\frac{t_e}{T_0}\right) G(t_e) = R, \quad (24)$$

where $R = C/B$ is the ratio of the initial chirp coefficient to the level of SPM experienced. The function $H_2(t/T_0)\,G(t)$ is plotted in Fig. 7(a) and has values comprised between $-2$ (attained at $t = 0$) and $4\exp(-3/2)$ (attained at $t_s = T_0\,(3/2)^{1/2}$). In other words, for $R$ below $R_{min} = -2$ or above $R_{max} = 4\exp(-3/2)$, the chirp $\delta\omega(t)$ is a strictly monotonic function as shown in panels b1 or b5 of Fig. 7 (blue curves). Hence, for this range of values of $R$, we should not expect spectral interference to occur and the spectrum should not show any oscillatory structure. When $R = C = 0$ (panel b3), we recover the case of an unchirped input pulse studied in Sec. 3. For $-R_{min} < R < 0$ (see panel b4 as an example), the chirp has a local minimum and a local maximum, with the maximum $\omega_{Cm} = \delta\omega(t_{Cm})$ located on the trailing edge of the pulse at

$$t_{Cm} = \frac{T_0}{2}\sqrt{2 - 4\,W_0\left(-\frac{R}{4}\exp\left(\frac{1}{2}\right)\right)}. \quad (25)$$

For $R = R_{min}$, the two extrema merge and the chirp has an inflexion point at $t = 0$ (panel b5, red curve). For $0 < R < R_{max}$, the chirp may feature a minimum and a maximum on the trailing edge of the pulse as illustrated in panel b2. The maximum point is given by Eq. (25), while the minimum value $\omega_{Cmin} = \delta\omega(t_{Cmin})$ occurs at

$$t_{Cmin} = \frac{T_0}{2}\sqrt{2 - 4\,W_{-1}\left(-\frac{R}{4}\exp\left(\frac{1}{2}\right)\right)}. \quad (26)$$

For $R = R_{max}$, the chirp exhibits an inflexion point at $t_s = t_{Cm} = t_{Cmin}$ (panel b1, red curve).

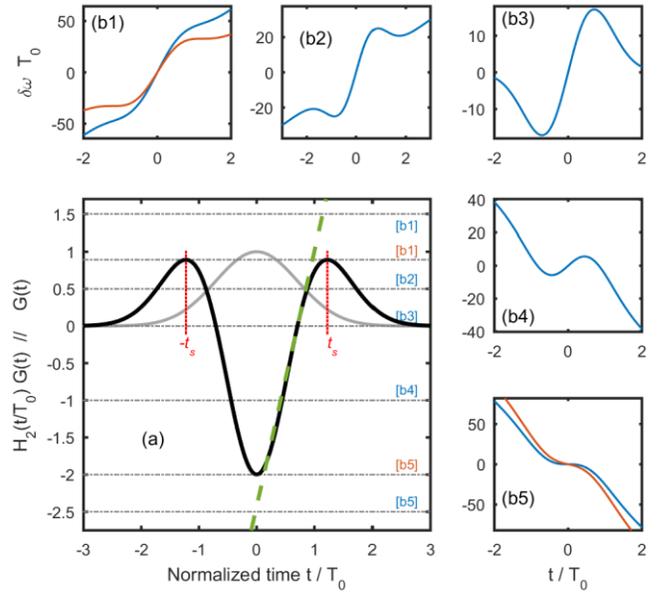

Fig. 7. (a) Plot of $H_2(t/T_0)G(t)$ for an initially chirped Gaussian pulse. The green dashed line is the linear approximation to the function in the vicinity of $t_m$ given by Eq. (26). The gray dash-dot horizontal lines represent the values of $R$ used in the different subplots b. (b) Temporal chirp profiles of the pulse for different values of $R$: (b1), $R = 1.5$ (blue curve) and $R = R_{max}$ (red curve); (b2) $R = 0.5$; (b3) $R = 0$; (b4) $R = -1$; (b5) $R = -2.5$ (blue curve) and $R = R_{min}$ (red curve).

The different behaviors of the chirp function according to the ratio of its linear and nonlinear parts impose remarkably different features on the pulse spectrum. To illustrate this point, we have plotted in Fig. 8 the evolution of the spectrum with the ratio $R$ at a level of SPM corresponding to $B = 20$ rad. The results highlight the distinctly different spectral regimes that are observed depending on $R$: for $R > R_{max}$, the central part of the spectrum does not show any oscillations. For $0 < R < R_{max}$, two intense peaks are visible in the spectrum and some ripple develops on the spectrum's boundaries. For $R_{min} < R < 0$, the spectrum features an oscillatory structure, which is different from the sinusoidal variation discussed in Sec. 3. Finally, for $R < R_{min}$, no oscillatory structure is visible in the spectrum and most of the pulse energy is focused on the center of the spectrum. Indeed, the pulse experiences spectral compression, a phenomenon that is well documented in the literature and has found many practical applications [5, 6, 29, 30].

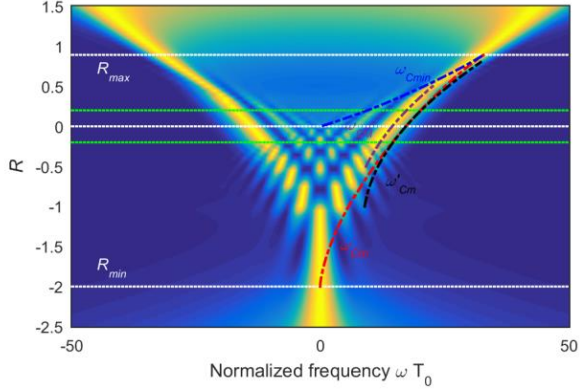

Fig. 8. Evolution of the spectrum of a chirped Gaussian pulse with the ratio $R = C/B$ for $B = 20$ rad, as obtained from numerical integration of Eq. (3). The white dotted horizontal lines represent the notable values of $R$. The green dotted horizontal lines indicate the $R$-values used in Figs. 9 and 11. The red, blue, black and purple dash-dotted curves represent the frequencies $\omega_{Cm}$, $\omega_{Cmin}$, $\omega'_{Cm}$ and $\omega'_{CM}$, respectively.

In the following subsection, we discuss some simplifications of the spectral interference model that can be advantageously used to describe the SPM-induced spectral pattern of an initially chirped Gaussian pulse. Then, we discuss the two regimes where the pulse spectrum exhibits an apparent oscillatory structure and explain the origin of the spectral features that are induced by the initial chirp of the pulse.

### B. Simplified model

The maximum point of the chirp $t_{Cm}$ given by Eq. (25) is very important for the description of the SPM pattern. However, once again the presence of the Lambert $W$ function would prevent the development of a fully analytical approach. We can circumvent this difficulty by considering an approximate formulation of the problem that is valid in the highly nonlinear propagation regime for which $R$ is close to zero. To this end, we approximate the function $H_2(t/T_0)G(t)$ around the point $t = t_m$ by its first-degree Taylor polynomial:

$$H_2\left(\frac{t}{T_0}\right) G(t) \simeq \frac{2^{\frac{5}{2}} e^{-\frac{1}{2}}}{T_0} (t - t_m). \quad (27)$$

The green dashed line in Fig. 7(a) highlights the validity of this approximation over a rather wide range of possible values of $R$. Accordingly, Eq. (25) simplifies to

$$t'_{Cm} = t_m + 2^{-\frac{5}{2}} e^{\frac{1}{2}} R\, T_0, \quad (28)$$

and the peak value of the chirp is obtained as

$$\omega'_{Cm} = \omega_m + \frac{C}{\sqrt{2}\, T_0}\left(1 + \frac{e^{\frac{1}{2}} R}{4}\right). \quad (29)$$

Equation (29) indicates that for $C > 0$, the maximum instantaneous frequency excursion contained in the pulse spectrum is increased relative to the case of an unchirped input pulse. Conversely, for slightly negative values of $R$, the maximum frequency is smaller than that for an unchirped pulse. The black dash-dotted curve in Fig. 8 confirms the validity of the approximate formula of Eq. (29) for small values of $R$.

To estimate the position of the outermost peak of the spectrum, we can use the method described in Sec. 3. To be rigorous, we should include in the total phase difference between pulse parts having the same instantaneous frequency a phase term accounting for the initial chirp of the pulse. However, we have found that for small values of $R$, this term does not affect the prediction of the position of the outermost peak $\omega'_{CM}$. Hence, for small $R$ we can utilize Eq. (20) with $\omega_m$ replaced by $\omega'_{Cm}$. The expression for $\omega'_{CM}$ so obtained is plotted in Fig. 8 with a purple dash-dotted curve and confirms the pertinence of the simplification made.

### C. Normal initial chirp

In this section, we focus on an input Gaussian pulse with a linear normal (positive) chirp such that the ratio $R$ is below $R_{max}$. The time-dependence of the total chirp of the pulse for $C = 4$ and $B = 20$ rad is shown in Fig. 9. We can clearly see that the chirp has a local maximum $\omega_{Cm}$ at $t_{Cm} < t_s$ and a local minimum $\omega_{Cmin}$ at $t_{Cmin} > t_s$. Accordingly, for a given instantaneous frequency $\omega_0$ such that $0 < \omega_0 < \omega_{Cmin}$, only one point in the pulse corresponds to this frequency, hence no spectral interference occurs. By contrast, for $\omega_{Cmin} < \omega_0 < \omega_{Cm}$, three instants: $t_1 < t_{Cm}$, $t_{Cm} < t_2 < t_{Cmin}$, and $t_3 > t_{Cmin}$ have instantaneous frequency equal to $\omega_0$. However, one should note that as $t_3$ is very far from the center of the pulse, the pulse intensity at this point $I_3 = G(t_3)$ is very low and can be neglected. This reduces the problem to a two-wave interference process similar to the one discussed in Sec. 3 for the case of an unchirped input pulse. As a result, spectral interference only occurs at the boundaries of the spectrum as can be seen in Fig. 10.

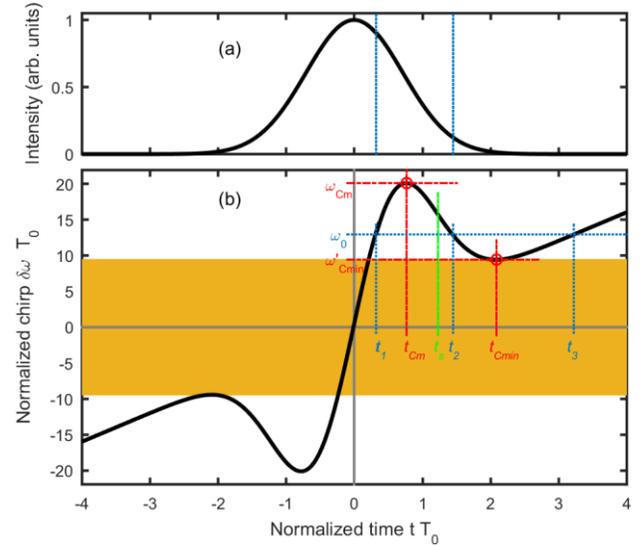

Fig. 9. Temporal intensity (panel a) and chirp (panel b) profiles for an initially chirped Gaussian pulse with $C = 4$ after undergoing SPM at a $B$-integral of 20 rad. The extreme points of the chirp function and the solutions of the equation $\delta\omega = \omega_0$ are indicated with red and blue lines, respectively.

When $R$ approaches $R_{max}$, $\omega_{Cmin}$ and $\omega_{Cm}$ become very close to each other. This results in the development of a single peak on each side of the spectrum, which thus acquires a 'batman-ear' shape. Accordingly, the use of a linearly chirped input pulse may be a convenient alternative to a triangular-shaped pulse for efficiently copying information onto two different frequency channels [31, 32]. In the opposite limit of $R$ approaching zero, which occurs for a low input chirp or for a high level of SPM accumulated in the fiber, the oscillations cover the major part of the spectrum.

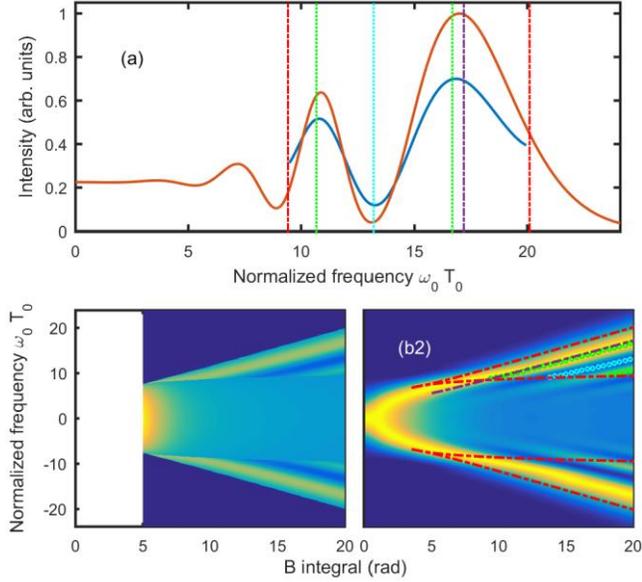

Fig. 10. SPM pattern of an initially chirped Gaussian pulse with $C = 4$. (a) Spectral intensity distribution $I(\omega)$ (blue solid curve) compared to the spectrum $S(\omega)$ obtained from Eq. (3) (red solid curve) for a $B$-integral accumulated in the fiber of 20 rad. (b) Evolutions of $I(\omega)$ (panel 1) and $S(\omega)$ (panel 2) with the $B$-integral. The red and purple vertical dash-dotted lines represent the extreme frequencies $\omega_{Cm}$ and $\omega_{Cmin}$, and the toy-model prediction for the position of the outermost spectral peak $\omega'_{CM}$, respectively. The green and cyan dotted lines or open circles represent the frequencies at which constructive and destructive interference occurs, as predicted by Eqs. (13) and (12), respectively.

conveniently reproduce the positions of the extrema of the oscillating pattern. However, as can also be seen in Fig. 12(b1), the details of the pattern are not qualitatively reflected. Indeed, $S(\omega)$ has a more complex structure than a sinusoidal-like variation, exhibiting an over-modulation and sharper features.

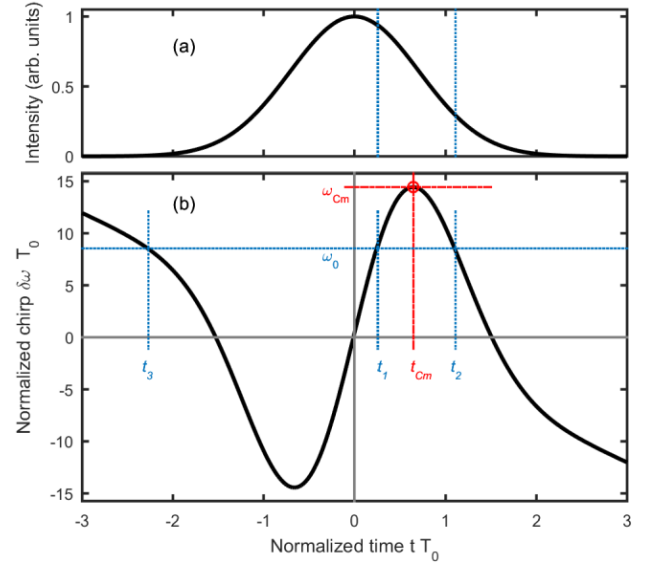

Fig. 11. Temporal intensity (panel a) and chirp (panel b) profiles for an initially chirped Gaussian pulse with $C = -4$ after undergoing SPM at a $B$-integral of 20 rad. The extreme points of the chirp function and the solutions of the equation $\delta\omega = \omega_0$ are indicated with red and blue lines, respectively.

To reproduce these properties, one has to take into account the third wave located at $t_3$. The resulting spectral intensity distribution is represented by a purple solid curve in Fig. 12(a) and is clearly in much better agreement with the actual spectrum than the two-wave interference pattern. The results shown in Fig. 12(b2) substantiate the necessity of including the third wave in the interference model for an accurate qualitative description of the spectral pattern engendered by SPM.

### D. Anomalous initial chirp

Finally, we discuss the case of an input Gaussian pulse with an anomalous (negative) chirp. As previously mentioned, for $R < R_{min}$ the pulse is spectrally compressed by the effects of SPM [5, 6, 29, 30]. We are interested here in values of $R$ above $R_{min}$, which correspond to the propagation regime in which the spectrum expands again after the point of maximum spectral focusing and, thus, spectral interference can arise. The temporal chirp profile of the pulse for $C = -4$ and $B = 20$ rad is shown in Fig. 11. Contrary to the case of a normal initial chirp, spectral interference can now occur for any value of $\omega_0$ such that $-\omega_{Cm} < \omega_0 < \omega_{Cm}$. For a given $\omega_0 > 0$, the pulse has this instantaneous frequency at three distinct points: $0 < t_1 < t_{Cm}$, $t_2 > t_{Cm}$, and $t_3 < -t_{Cm}$.

In Fig. 12(a) we compare the spectrum $I(\omega)$ reconstructed from the interference between $t_1$ and $t_2$ with the actual spectrum $S(\omega)$ obtained from Eq. (3). We can see that the two-wave interference model can

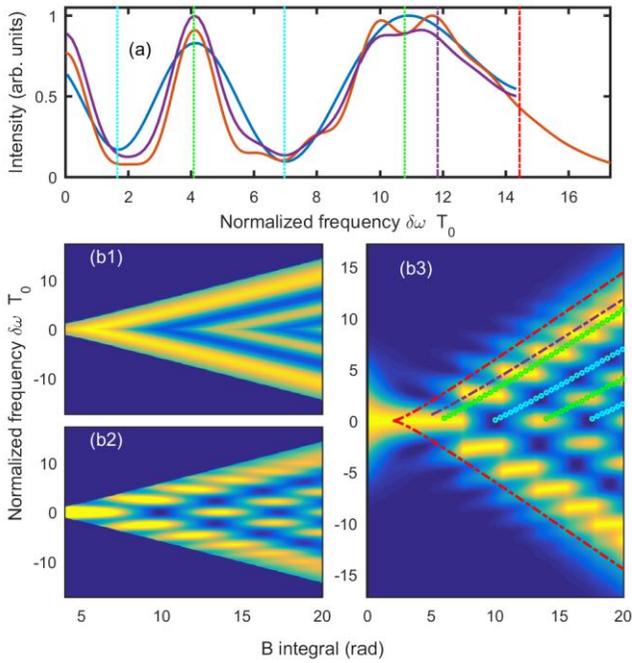

Fig. 12. SPM pattern of an initially chirped Gaussian pulse with $C = -4$. (a) Spectral intensity distribution $I(\omega)$ resulting from the interference of two and three waves (blue and purple solid curves respectively) compared to the spectrum $S(\omega)$ obtained from Eq. (3) (red solid curve) for a $B$-integral accumulated in the fiber of 20 rad. (b) Evolutions of $I(\omega)$ when two and three interfering waves are considered (panels 1 and 2, respectively) and $S(\omega)$ (panel 3) with the $B$-integral. The red and purple vertical dash-dotted lines represent the maximum frequency $\omega_{Cm}$ and the toy-model prediction for the position of the outermost spectral peak $\omega'_{CM}$, respectively. The green and cyan dotted lines or open circles represent the frequencies at which constructive and destructive interference occurs, as predicted by Eqs. (13) and (12), respectively.

## 6. CONCLUSIONS

We have described a simple theoretical approach to predict the main features of the spectra of optical pulses affected by SPM, which is based on regarding the optical spectrum modification as an interference effect. We have shown that a two-wave interference process is sufficient to describe the SPM-broadened spectra of initially Fourier-transform-limited pulses or pulses with an initial positive linear chirp, and to accurately predict the extreme values of the spectra. Simplified but fully tractable closed formulae have been derived for the positions of the outermost peaks of the spectra, which provide a more plausible measure of the spectrum extent than the approximate or rms expressions for the spectral bandwidth that are commonly used. We have also discussed the various spectral regimes that are observed depending on the ratio of the initial chirp of the pulses and the level of SPM accumulated in the fiber, and shown that in the case of negatively chirped input pulses, the description of the SPM spectral patterns requires the inclusion of a third wave in the interference model.

The present approach has been discussed in the context of dispersionless passive propagation, but its extension to include gain or loss is straightforward. Our approach can also help better understand qualitatively the genesis of peculiar spectrum shapes of laser pulses, such as the batman ear spectra observed in all-normal dispersion lasers [33] or Mamyshev oscillators [34]. While the present discussion focuses on SPM, the concept can also be applied to other nonlinear modulations of the phase of a pulse, such as the modulation generated by cross-phase modulation [35] or an external modulator [36-38].

**Funding Information.** Institut Universitaire de France (IUF).

**Acknowledgment.** We thank J. Fatome, B. Kibler and K. Hammani for fruitful discussions.